\documentclass[11pt]{article}
\usepackage{moriond,epsfig}

\def\be{\begin{equation}}
\def\ee{\end{equation}}
\def\bea{\begin{eqnarray}}
\def\eea{\end{eqnarray}}

\usepackage{subfigure}
\begin{document}
\vspace*{4cm}
\title{Searches for new phenomena at CMS and ATLAS}

\author{ Tanja Rommerskirchen (for the CMS and ATLAS collaborations) }

\address{Physics Institute, University of Zurich, Switzerland}

\maketitle\abstracts{The prospects of the ATLAS and CMS experiments at LHC for beyond standard model searches are described. These studies concentrate on the search plans for supersymmetry (SUSY) and beyond in the first few years of data taking.}

\section{Introduction}
The first collisions at the  Large Hadron Collider are foreseen for the end of 2009. This article summarizes the plans of the ATLAS and the CMS collaborations for beyond standard model searches with early LHC data.
The collision energy assumed in  the presented searches is $\sqrt{s} = 14$ TeV.
The article is divided into two parts, one part dealing with supersymmetry (SUSY) as theoretically preferred candidate for new physics; the other part presents plans on how to ensure sensitivity for even more exotic signatures.

\section{SUSY searches}\label{sec:SUSYsearches}
 Even the supersymmetric model with the smallest number of additional particles, the MSSM (Minimal Supersymmetric Model),  is characterized by more than 100 parameters \cite{PTDR}. 
  In order to cover as much supersymmetric parameter space as possible, searches concentrate on common features of most supersymmetric models, such as multiple jets plus leptons and missing energy for R-parity conserving models. 
  Accordingly, SUSY search channels are defined in terms of the expected event topology, \it{e.g.} \rm number of jets, leptons and photons.
  
  \subsection{Search channels in mSuGra} 
  Two nice example for this kind of searches are the  same-sign di-lepton search prepared by the ATLAS collaboration \cite{CSC} and the two jets plus zero lepton search prepared by the CMS \cite{dijet} collaboration.
 
  \begin{figure}[htp]
 \begin{center}
 	\subfigure[Distribution of $E_T^{miss}$ in the di-lepton channel. A cut on $E_T^{miss} >$ 100~GeV effectively suppresses most of the background, leaving the shaded area.]{\label{fig:dilepton}\includegraphics[scale=0.4]{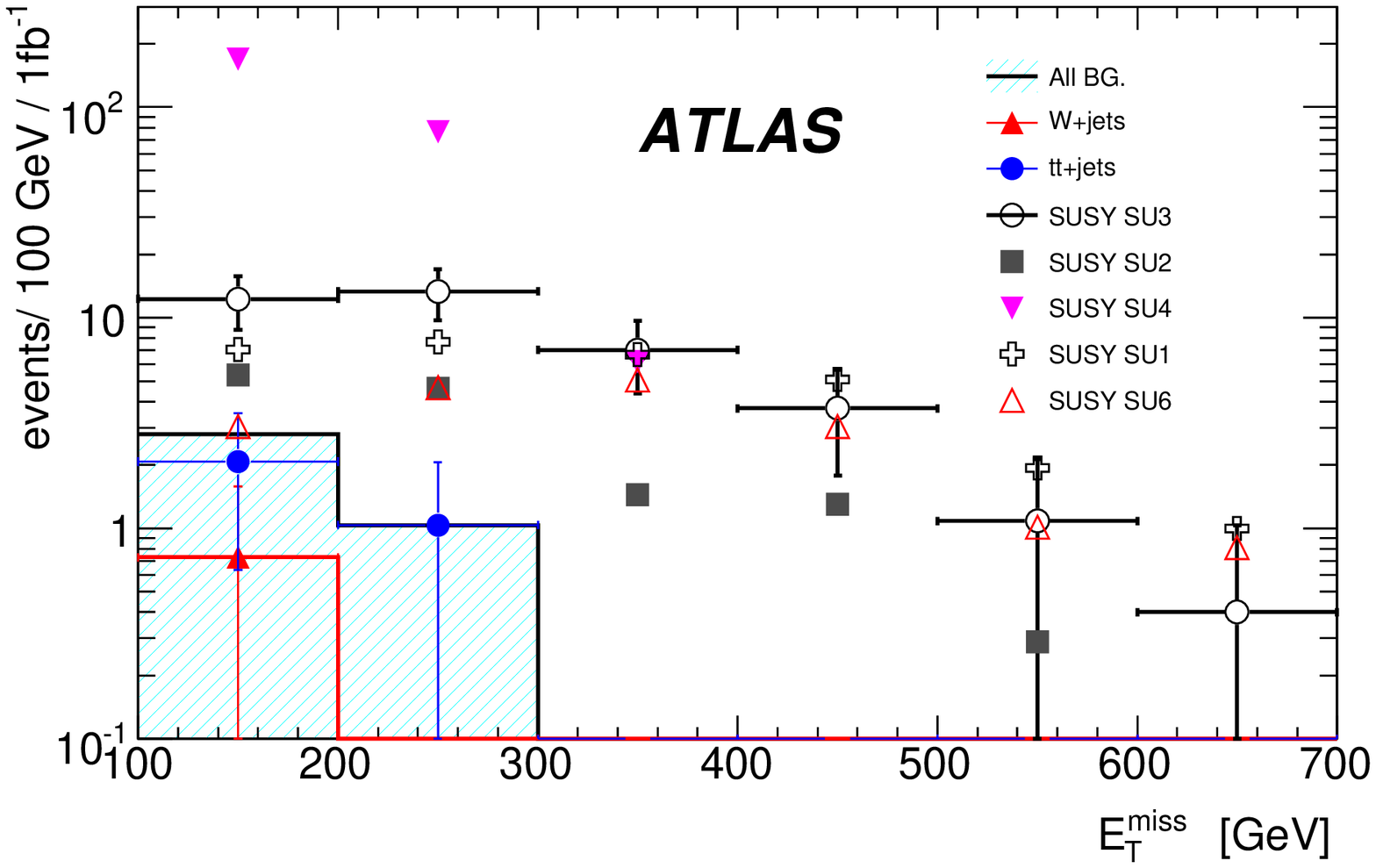}}
	\subfigure[Distribution of $\alpha_T$ in the di-jet channel.]{\label{fig:alphaT}\includegraphics[scale=0.7]{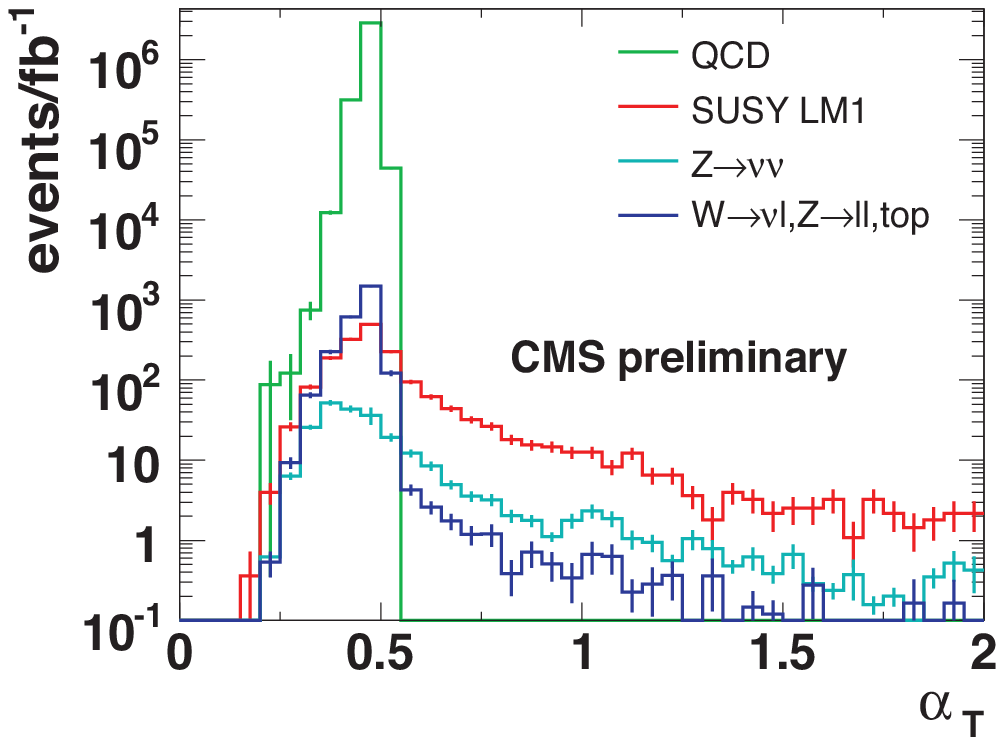}}
   \end{center} 
\caption{
\label{fig:SUSYdiscovery}$\alpha_T$ and $E_T^{miss}$ distributions shown for 1~\rm{fb}$^{-1}$ of integrated data.} 
\end{figure}
 
  The basic requirements in the same-sign di-lepton search are two same-sign di-leptons plus four jets plus missing transverse energy ($E_T^{miss}$) larger than 100 GeV. Figure \ref{fig:dilepton} shows the $E_T^{miss}$ distribution for several points in the mSuGra parameter-space and standard model background. After a selection, nearly no standard model background is left. During the early period of data taking the usage of  calorimeter based $E_T^{miss}$  might be problematic. From experience at the Tevatron, large systematic uncertainties are expected.
This uncertainty is reduced in the two jet + zero lepton + $E_T^{miss}$ search, which profits from the particular event topology by defining a  non-$E_T^{miss}$ based variable
\begin{equation}
\alpha_T=\frac{E_T^{j2}}{M_{inv,T}^{j1,j2}},
\end{equation}
where $E_T^{j2}$ is the energy of the second leading jet and $M_{inv,T}^{j1,j2}$ is the transverse invariant mass of the two jets.
$\alpha_T$ is a powerful variable which discriminates between events containing real missing energy in form of neutrinos or neutralinos and the standard model di-jet background. This quantity is also particularly robust against jet-energy mismeasurements. Figure \ref{fig:alphaT} shows the $\alpha_T$-distribution for standard model backgrounds and for one point in the mSuGra parameter-space.

\subsection{Background estimation methods}
Because searching for supersymmetry means searching for the unknown, to avoid claiming  a false discovery, understanding and controlling the standard model background is crucial. In addition to the usual Monte Carlo based simulations, data driven methods have been developed by both experiments.
   Especially for the non-reducible $Z\rightarrow\nu\bar{\nu}$ background, several complementary estimation methods are in place.
  The general idea of these methods is to select a clean sample for the  control background which is similar to $Z\rightarrow\nu\bar{\nu}$ and to correct for the differences in e.g. cross-section and acceptance between the control sample and $Z\rightarrow\nu\bar{\nu}$. 
 Control backgrounds for $Z\rightarrow\nu\bar{\nu}$ that have been studied are $Z\rightarrow ll$ \cite{CSC}, photon + jets \cite{Zinvisible}, and $W\rightarrow l\nu$ \cite{dijet}.

\subsection{SUSY discovery reach}
Most of the searches presented so far use simulations for particular points in the supersymmetric parameter-space to study the kinematics expected in real events. These points have been chosen by both collaborations and should cover most of the possible topologies in a supersymmetric model. Complementary to these points, scans in the supersymmetric parameter-space have been made in order to estimate the discovery reach of standard SUSY {\rm $E_T^{miss}$ + $m$ jets + $n$ lepton} searches in various models.  In Figure \ref{fig:SUSYdiscovery} the estimated discovery reach of mSuGra searches done with ATLAS in the $n$ jet plus 0 lepton channel is illustrated \cite{CSC}.  Comparable studies  for the CMS detector have shown that both experiments have similar discovery reaches \cite{PTDR}.

 \begin{figure}
 \begin{center}
    \includegraphics[width=0.5\linewidth]{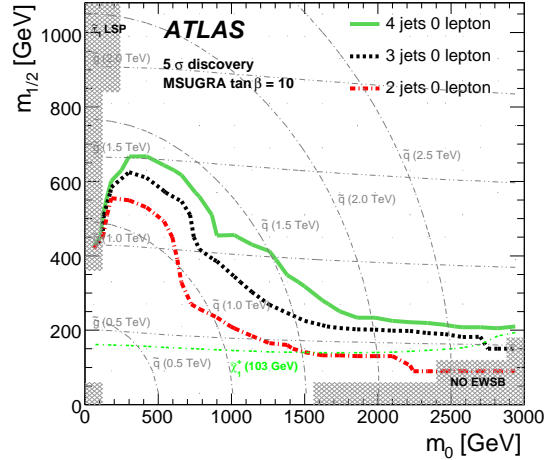}
 \end{center} 
 \caption{
\label{fig:SUSYdiscovery}ATLAS discovery reach for the n jet plus 0 lepton channel in the $m_0$ and $m_{1/2}$ mSuGra plane for $\tan\beta=10$ at 1 ~fb$^{-1}$} 
\end{figure}

\section{Searches beyond mSuGra}\label{sec:bmSuGra}
 In addition to the standard mSuGra searches, both experiments aim for being sensitive to various unusual signatures predicted by SUSY models beyond mSuGra and by non-SUSY models.
  
  \subsection{Heavy stable charged particles}
\begin{figure}
 \begin{center}
    \includegraphics[width=0.4\linewidth]{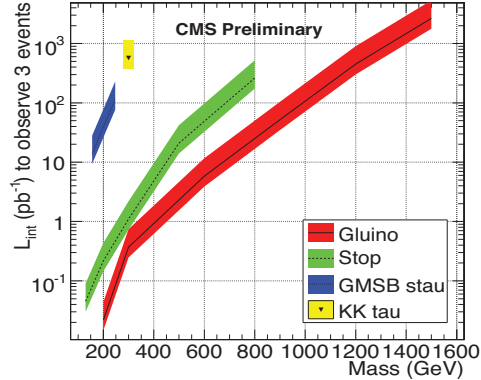}
 \end{center} 
 \caption{
\label{fig:HSCP}Integrated luminosity needed for discovering three HSCP (Heavy stable charged particle) events in the case that the full selection as described in ref. 5)  is applied and no standard model background is left. Four different models are tested.} 
\end{figure}

  Particularly challenging signatures are produced by heavy stable charged particles (HSCP), such as those predicted by supersymmetric models like GMSB and split SUSY, and non-supersymmeteric models such as universal extra dimensions (UED). HSCP could also be states composed of gluinos and stops. These states are called R-Hadrons. R-Hadrons are hadronically interacting with matter. The quarks bound to the stops or gluinos can be emitted and therefore the total charge of the hadron is modified. 
  Slow heavy particles are  difficult to trigger on as they might arrive too late in the muon system and be lost. Also the muon reconstruction in general would be non-trivial for charge-flipping R-Hadrons.
 Two independent measurements of the particle velocity, one in the muon chambers and one in the silicon tracker, provide a way to effectively suppress all standard model background. Figure \ref{fig:HSCP} shows the integrated luminosity needed by CMS to discover more than three HSCP events, for four different models, after the full event selection \cite{HSCP}. 
 Methods to make the trigger and pattern recognition more sensitive to this unusual signature are currently studied.
  
  \subsection{Di-object signature searches}
  Di-object searches are \it{e.g.}, \rm di-lepton, di-jet and di-photon searches which are sensitive to  new signals like heavy mass resonances as predicted by GUT, extra dimension and compositness models. 
  Di-object searches are also a simple way to address a broad class of new physics with small statistics.  Also lepton + $E_T^{miss}$ and jet + $E_T^{miss}$ searches are being prepared to cover heavy W-like bosons as predicted by the left-right model and mono-jet final states predicted by extra dimension models.
Among these different channels, di-muon \cite{highMassResonancesMuon} and di-electron resonances \cite{highMassResonancesEl}  are certainly the golden channels for early searches due to their clean signatures.
 
 \subsection{More exotic searches}
Some of the prepared searches are not aimed at early data but cover interesting signatures which would become visible  with higher luminosities and a better understanding of the detector.  This includes searches for vector boson resonances \cite{CSC}, fourth generation quarks \cite{fourthgen} and mini black holes \cite{CSC}.
In order to complement this searches for specific signatures a model unspecific search (MUSIC) \cite{MUSIC} has been set up by the CMS collaboration. MUSIC automatically sorts the incoming events according to their signature, e.g. number of jets, leptons and tests the number of events against the standard model background estimation. For early data MUSIC is expected to improve the understanding of the detector response and the MC simulations.


\section*{References}
\begin{thebibliography}{99}
\bibitem{PTDR}CMS collaboration, {\em CMS PTDR 2}, CERN/LHCC 2006-021.
\bibitem{CSC}ATLAS collaboration, {\em Expected Performance of the ATLAS Experiment, Detector, Trigger and Physics}, CERN-OPEN-2008-020, Geneva, 2008.
\bibitem{dijet}CMS collaboration, {\em SUSY searches with di-jet events}, CMS PAS SUS-08-005.
\bibitem{Zinvisible}CMS collaboration, {\em Data driven estimation of invisible Z background, in the SUSY MET plus jets search}, CMS PAS SUS-08-002.
\bibitem{HSCP}CMS collaboration, {\em Search for heavy stable charged particles with 100pb-1 and 1fb-1 in the CMS experiment}, CMS PAS EXO-08-003.
\bibitem{highMassResonancesMuon}CMS collaboration, {\em Search for new high-mass resonances decaying to muon pairs in the CMS experiment}, CMS PAS SBM-07-002.
\bibitem{highMassResonancesEl}CMS collaboration, {\em Search for high-mass resonances decaying into an electron pair in the CMS experiment}, CMS PAS EXO-08-001.
\bibitem{Wprime}CMS collaboration, {\em Discovery potential of W'$\rightarrow$e$\nu$ at CMS}, CMS PAS EXO-08-004.
\bibitem{dijetExo}CMS collaboration, {\em CMS search plans and sensitivity to new physics using dijets}, CMS PAS SBM-07-001.
\bibitem{monojet}CMS collaboration, {\em Search for mono-jet final states from ADD extra dimensions}, CMS PAS EXP-08-011.
\bibitem{fourthgen}CMS collaboration, {\em Search for heavy bottom-like fourth generation quark pair production at CMS in pp collisions at $\sqrt{s}=14$ TeV}, CMS PAS EXO-08-009.
\bibitem{MUSIC}CMS collaboration, {\em MUSIC-an automated scan for deviations between data and Monte Carlo simulation}, CMS PAS EXO-0-005.



\end{thebibliography}
\end{document}